\documentclass[journal]{IEEEtran}
\usepackage{cite}
\usepackage{amsmath,amssymb,amsfonts}
\usepackage{graphicx}
\begin{document}

\title{On the use of GNSS for Automatic Detection of Attenuating Environments}

\author{\IEEEauthorblockN{Stefan Ruehrup}

\IEEEauthorblockA{
\textit{ITS Services} \\
\textit{ASFINAG Maut Service GmbH}\\
Vienna, Austria } \\
}

\maketitle

\begin{abstract}
When different radio applications share the same spectrum, the separation by attenuating material is a way to mitigate potential interference. The indoor restriction for WLAN devices in 5150-5350 MHz is an example for a regulatory measure that aims at having WLAN devices operating in an environment that provides sufficient attenuation to enable sharing with other services \cite{eccdec0408}.

In this paper we investigate whether an attenuating environment can be automatically detected without user interaction. Instead of detecting an indoor location, we are directly looking for a detection of an attenuating environment.

The basic idea is that signals from global navigation satellite services (GNSS) can be received practically everywhere on earth where there is a view to the sky. Where these signals are attenuated, the receiving device is assumed to be in an attenuating environment. In order to characterize such environment, we evaluate the detectable GNSS satellites and their carrier-to-noise density. Example measurements show that GNSS raw data can help to distinguish between low-attenuating  locations and higher-attenuating locations. These measurements were conducted with GNSS receiver in an off-the-shelf Android tablet in order to show the feasibility of the approach. 
\end{abstract}


\section{Introduction}

WLAN devices use frequency bands assigned to Wireless Access Systems and Radio Local Area Networks (WAS/RLAN). These frequency bands are shared with other radio services or applications, and certain mitigation techniques are required by regulation. The indoor restriction for WLAN devices in 5150-5350 MHz is an example for a regulatory measure that aims at allowing WLAN devices in environments where WLAN signals are attenuated. A definition of \emph{indoor use} is given in \cite{eccdec0408}: 
\begin{quote}
Indoor use is intended to mean inside a permanent domestic or commercial building which will typically provide the necessary attenuation to facilitate sharing with other services.
\end{quote} 
While it is easy for users to check whether their WLAN devices are installed or used inside a building, it is hard to specify the amount of attenuation the building provides to the outside world. The term ``indoor'' in this context can be seen as an approximation of an \emph{attenuating environment}. There are locations inside buildings that do not provide significant attenuation, and there are attenuating environments, which are not inside a building. Therefore, we do not aim at detecting an indoor location, but to detect an attenuating environment. The goal is to perform this detection automatically, in order to relieve users from questions whether a building with windows and walls has certain attenuating properties. 

\begin{figure}[htbp]
\centerline{\includegraphics[width=\linewidth]{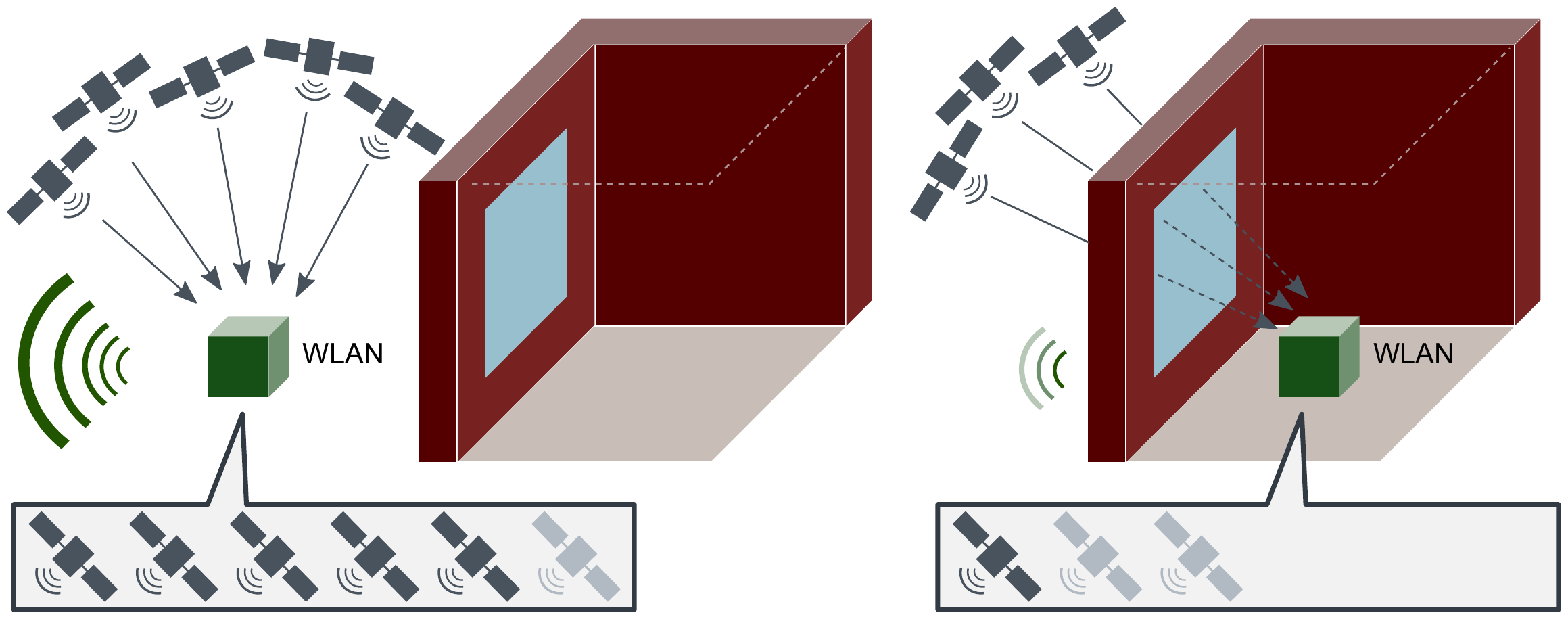}}
\caption{In an attenuating environment (right) the WLAN device receives attenuated GNSS signals. At the same time its own signals are also attenuated towards the outside of the environment.}
\label{fig:attenuating}
\end{figure}

The main idea is to measure the signals from global navigation satellite services (GNSS) and their attenuation at the device under consideration, e.g. a WLAN device with GNSS receiver. If the GNSS signals are significantly attenuated, then also the devices' emitted WLAN signals are assumed to be attenuated (see Figure~\ref{fig:attenuating}). 

The unique feature of global navigation satellite systems (GNSS) is their availability in most parts of the world. Satellites are traversing the open sky in practically all parts of the world -- except for polar regions, where above a certain latitude the satellites do not exceed a certain elevation \cite{dejong2014precise}. Satellites of the GNSS-Systems GPS, GLONASS, Galileo, Beidou, and QZSS are emitting signals in the frequency ranges 1164-1215 MHz, 1215-1300 MHz and 1559-1610 MHz. GNSS receivers can be found in smart phones, tablets, watches, machinery and many other types of equipment. It is already possible to extract raw data and can deliver the visible satellites as well as their carrier-to-noise density. The carrier-to-noise density ($C/N_0$) is of specific interest, because it is an indicator of the signal quality that is 
independent of the receiver's acquisition algorithm \cite{joseph2010gnss}. 

This paper investigates how these GNSS measurements can be used to distinguish between attenuating environments and non-attenuating environments.

This paper is structured as follows:
In Section~\ref{sec:related} we review related work. 
In Section~\ref{sec:measurement} we describe our measurement setup. 
In Section~\ref{sec:results} we present the measurement results. 
In Section~\ref{sec:detection} we derive criteria for detection of attenuating environments. 
In Section~\ref{sec:discussion} we discuss aspects for generalisation. 
Section~\ref{sec:conclusion} concludes the paper.

\section{Related Work} \label{sec:related}

The detection of attenuating environments is closely related to indoor detection. 
The detection of indoor locations, especially by mobile phone apps, has been intensively studied. Not only GNSS-based solutions, but also other sensor inputs have been considered. As an example, Li et al. \cite{li2014iodetector} proposed a combination of light sensors, magnetism sensors, and cell tower signals. Detailed explanations are given for each of the data sources. Though we are not aiming to detect indoor locations, the approaches for indoor detection could be candidates for the detection of an attenuating environment. 

In \cite{chen2017satprobe}, GNSS is specifically used for indoor detection. The authors show that by evaluating the number of visible satellites, the environment can be classified as indoor, outdoor, or semi-outdoor. The evaluation of the number of satellites was chosen for energy-efficiency reasons. The authors propose further refinements in signal processing and computation to reduce energy consumption. In this paper, we do not use dedicated hardware, but try to base our proof of concept on a mass market device. The implementation in this paper uses the Android API for raw GNSS measurements, which  is comprehensively described in \cite{gsa17whitepaper}. Further applications based on more advanced chipsets are described in\cite{fortunato2019real-time}, including a characterisation of the measured signals.

Attenuation measurements have been performed for building materials and vehicles. In \cite{angskog2015measurement} measurements for window attenuation in the range of 1 Ð- 18\,GHz are presented, which show a huge span of attenuation values across several window types, especially when diffenent types of glass coating are compared. The penetration of GNSS signals though building materials has been characterised in \cite{hein2008gnss} and \cite{kjaergaard2010indoor}.
In \cite{schmidt08uwb} measurements in the range of 2-8 GHz were conducted for an aircraft. These measurements could be regarded as a characterisation whether an aircraft is an attenuating environment. The measurements were not aiming at the detection of the attenuating environment, but to quantify the attenuation provided by the hull and structure of the aircraft. 

\section{Measurement Setup} \label{sec:measurement}

\subsection{Hardware}
Measurements were performed with a Samsung Galaxy Tab S2 (model number SM-T715), which was introduced in 2015 and supports GPS and GLONASS. It was the intention not to use specialized hardware or high-end GNSS chipsets, but to demonstrate the ability of the approach on consumer hardware that has been available for several years and became widely available. It should however be noted, that improvements in GNSS chipsets could deliver more precise measurements through their support of more GNSS systems or dual frequency receivers. The tablet serves only as a platform for GNSS measurements, which was chosen because it allows to conveniently take measurements in several locations -- a GNSS chip could also be built into other devices, such as access points.

\subsection{Software}
Measurements were recorded with the Android GnssLogger\footnote{https://github.com/google/gps-measurement-tools/releases/tag/2.0.0.1}. The GnssLogger is an open source application that demonstrates the access to the GNSS raw data API functions introduced in Android 7.0. For our measurements, the app was modified such that it records the number of visible satellites and the carrier-to-noise per satellite via the \texttt{GnssStatus}\footnote{https://developer.android.com/reference/android/location/GnssStatus} object rather than then \texttt{GnssMeasurement} object. In comparison to \texttt{GnssStatus}, the \texttt{GnssMeasurement} object provides further data elements that are only supported by a smaller number of chipsets, but are not needed for our study. See Table~\ref{tab:gnssstatus} for an overview of the methods provided by \texttt{GnssStatus} to access raw data. 

The software uses only API calls that were available in Android 7.0 (API level~24)\footnote{see https://developer.android.com/guide/topics/sensors/gnss}, which is the earliest version when GNSS raw data was made accessible. Callbacks to \texttt{GnssStatus} are triggered roughly every second, such that maeasurements were recorded in 1-second intervals.

\begin{table}[htbp]
\caption{Methods of GnssStatus in Android API Level 24 (excerpt)}
\begin{center}
\begin{tabular}{|l|l|}
\hline
Aggregated:
 & getSatelliteCount ()  \\
\hline
Per Satellite:
 & getAzimuthDegrees (int satelliteIndex) \\
 & getCn0DbHz (int satelliteIndex) \\
 & getConstellationType (int satelliteIndex) \\
 & getElevationDegrees (int satelliteIndex) \\
 & getSvid (int satelliteIndex) \\
 & hasAlmanacData (int satelliteIndex) \\
 & hasEphemerisData (int satelliteIndex) \\
 & usedInFix (int satelliteIndex) \\
\hline
\end{tabular}
\label{tab:gnssstatus}
\end{center}
\end{table}

\subsection{Locations}

\begin{figure}[htbp]
\centerline{\includegraphics[width=\linewidth]{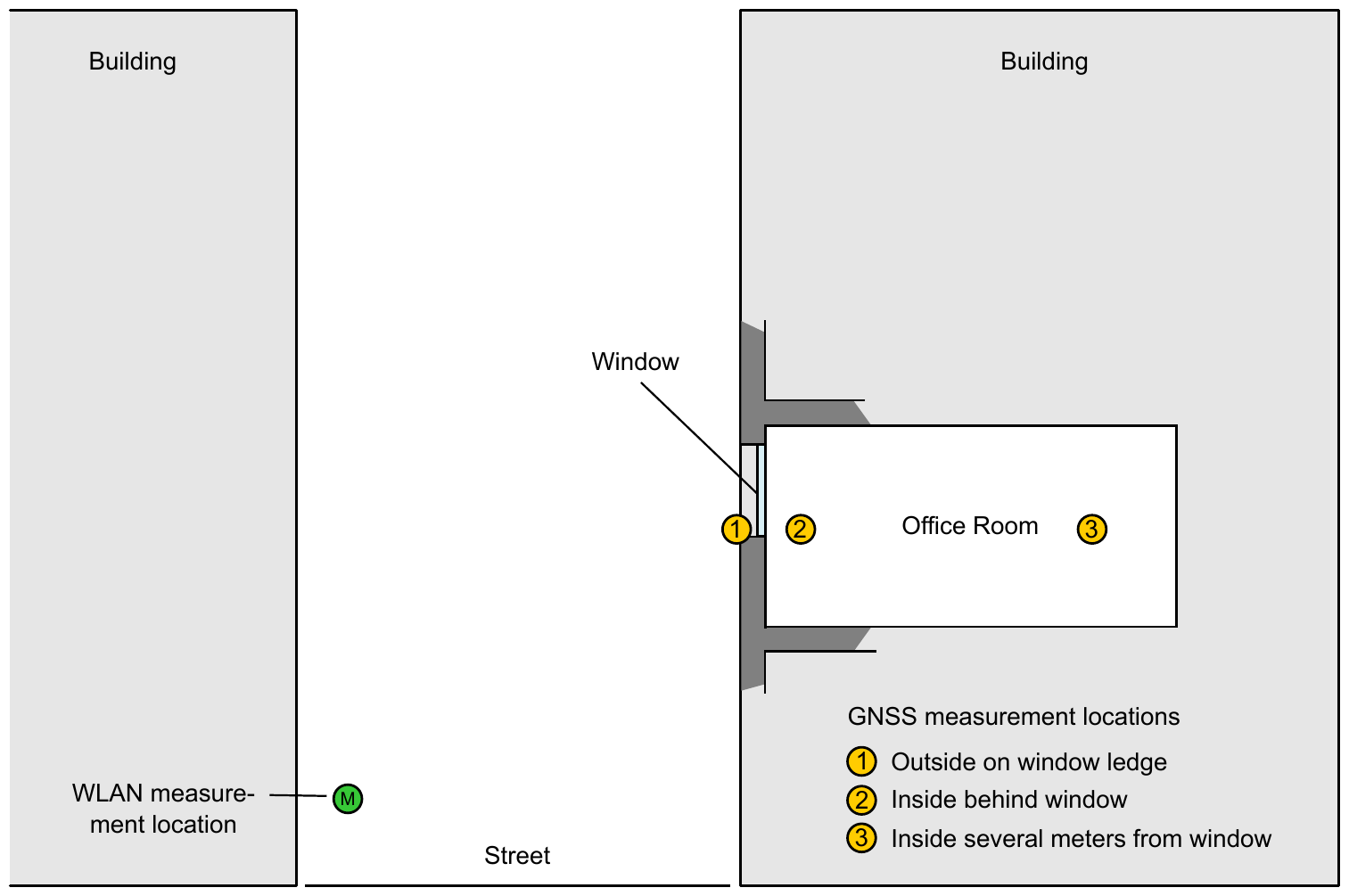}}
\caption{Office and urban canyon environment. 
GNSS measurements were taken at location 1, 2, and 3. 
Signals from a 5GHz WLAN inside the office room were measured outside at location M}
\label{fig:measurement-setup}
\end{figure}

Measurements were taken in urban and rural environments.
Urban measurements were taken in an office building as sketched in Figure~\ref{fig:measurement-setup}. 
Location 1 is located outdoor on the window ledge with a view to the sky, which is partially obstructed by the buildings. 
Location 2 is inside the building approximately 0.5\,m away from a row of closed windows at a height of approximately 0.8\,m from the ground. The windows have plastic frames and plastic panels in between.
Location 3 is approximately 4\,m from the closed window, where the sky is not visible any more, not even partially. The room has 4 windows with a size of 1\,m$\times$0.75\,m each. The distance between buildings is 16\,m and the building height is over 20\,m. Buildings have several rows of windows that are potential reflectors. 

Further urban measurements were taken on a nearby parking lot with very good visibility of the open sky. 

In rural area, measurements were taken on a hilltop, in the forest, and inside a waiting booth on a commuter railway station. The waiting booth has glass on all sides from floor to the ceiling, including a glass door. The glass had no visible coating, nor insulation.

A comparison measurements of the attenuation of a 5\,GHz WLAN signal was taken with a professional signal analyzer. A WLAN access point was placed inside the room close to the window with direct line of sight to the WLAN measurement location outside.

\subsection{Identification of attenuating environments}
In order to identify whether the office room is an attenuating environment, a 5GHz WLAN access point with 3 omnidirectional antennas was tuned to a 20 MHz channel centered at 5240 MHz. The signals were received outside on the other side of the street. A professional spectrum analyzer and a directional antenna pointed towards the window was used to record the maximum channel power over 20 MHz.  Direct line of sight to the WLAN access point at the open window was considered as reference.

The results in Table~\ref{tab:attenuation} show a significant attenuation. 
Therefore, Locations 2 and 3 are considered in an attenuating environment.

\begin{table}[htbp]
\caption{Attenuation measurements for the office environment}
\label{tab:attenuation}
\begin{center}
\begin{tabular}{lc}
Location & Measured attenuation \\
\hline
WLAN behind open window 	& reference   \\
WLAN behind closed window 	& -15 dB	  \\
WLAN in room (closed window) & -16 	dB	 \\
\hline \\
\end{tabular}
\end{center}
\label{default}
\end{table}

The window attenuation was measured separately with a WLAN access point using a directional patch antenna. It was placed directly at the window at a distance of 1-2\,cm, pointing through the window in the direction of the receive antenna of the spectrum analyzer, which was placed 4\,m apart. Measurements were taken with and without window. The comparison showed an attenuation in the range of 20 to 25 dBm. This result and a slightly visible tint are an indication of coated or metalized glass, which is often used in office buildings. Uncoated single layer glass showed an attenuation of 7 dBm. Low attenuating material was found in the plastic frames and panels between the windows, which explains the overall attenuation being less than the window attenuation.

The rural locations are not considered an attenuating environment.

\section{GNSS Measurement Results} \label{sec:results}

The measurement results for each location are presented in a pair of plots, where the upper plot shows the carrier-to-noise ($C/N_0$) values for each satellite. The lower plot shows the number of satellites from which signals were received, as well as the number of satellites use in a localization solution (``fix''). The number of satellites used in the fix is 0 if the location cannot be determined.

\begin{figure*}[htbp]
\centerline{\includegraphics[width=.9\linewidth]{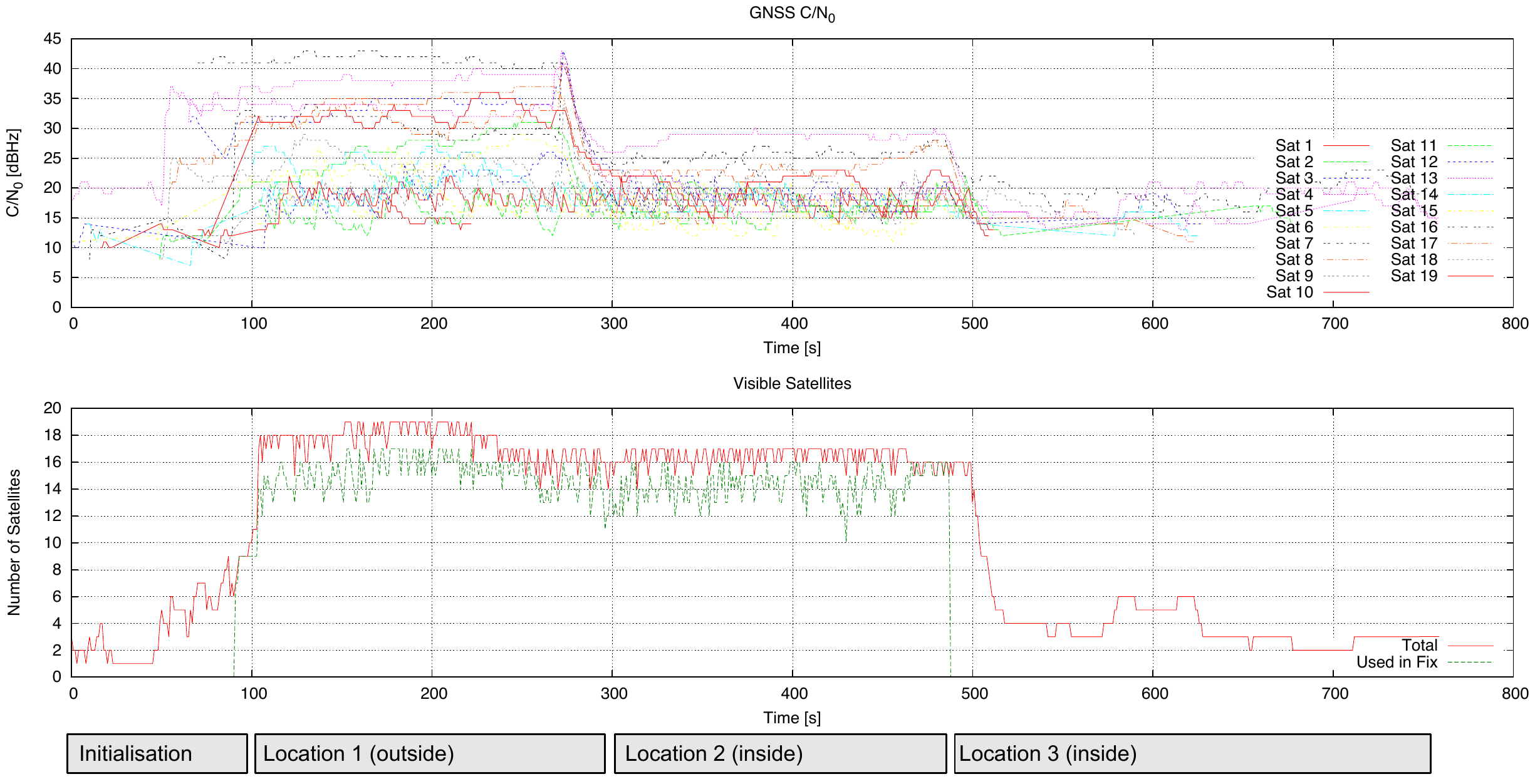}}
\caption{Measurement where the device is moved between locations 1, 2 and 3 in the office and urban canyon environment. See Figure~\ref{fig:measurement-setup} for a description of the locations.}
\label{fig:measurement-moving}
\end{figure*}

The measurement in Figure~\ref{fig:measurement-moving} shows a measurement where the device is moved between locations 1, 2 and 3 with a certain stable time between these locations. The number of satellites, including those used for the fix (lower plot) is similar for location 1 (outside) and location 2 (inside). Both situation cannot be clearly distinguished when using only the number of satellites. In the upper plot, however, one can observe a difference between outdoor (1) and indoor locations (2,3): in the outdoor location there is a relevant fraction of satellites with a $C/N_0$ exceeding 30\,dB-Hz, while this is not the case for the indoor location behind the attenuating walls and windows. This shows that $C/N_0$ values measured per satellite are relevant for characterizing an attenuating environment.

Note that these measurements should not be used to derive universal thresholds, since values depend on chipset, antenna and need to be calibrated.

\begin{figure*}[htbp]
\centerline{\includegraphics[width=.9\linewidth]{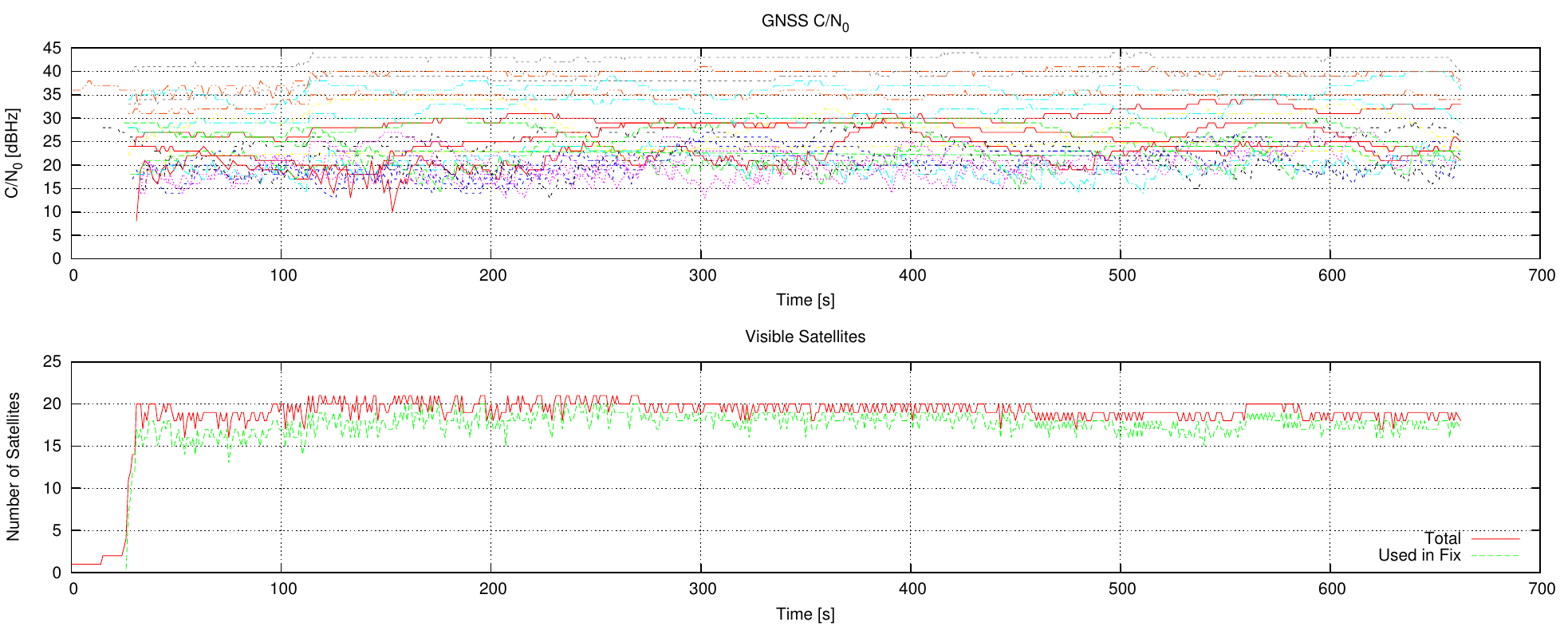}}
\caption{Example measurement at Location 1 (urban canyon) outdoor in front of the window. The device is initialised at this location and stays there during the measurement.}
\label{fig:measurement-location-1-only}
\end{figure*}

\begin{figure*}[htbp]
\centerline{\includegraphics[width=.9\linewidth]{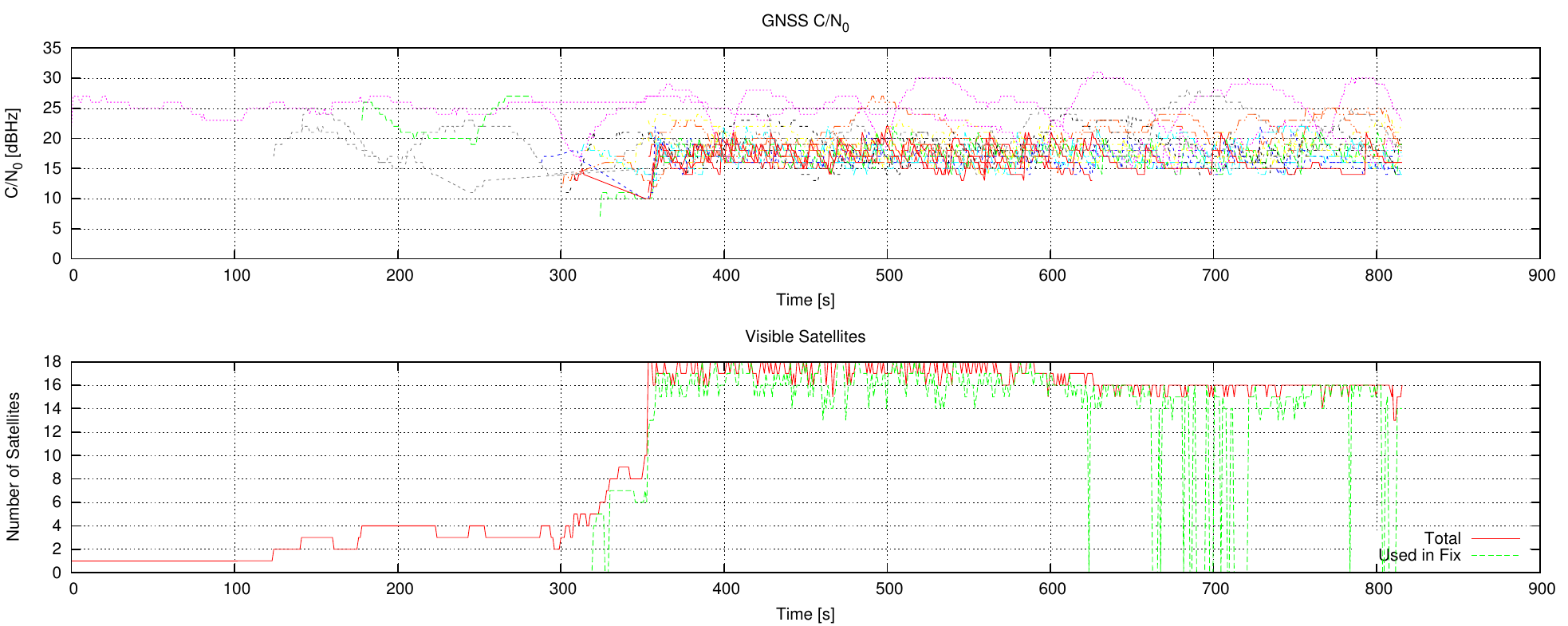}}
\caption{Example measurement at Location 2 (office) indoor behind the window with partial view to the sky. The device is initialised at this location and stays there during the measurement.}
\label{fig:measurement-location-2-only}
\end{figure*}

\begin{figure*}[htbp]
\centerline{\includegraphics[width=.9\linewidth]{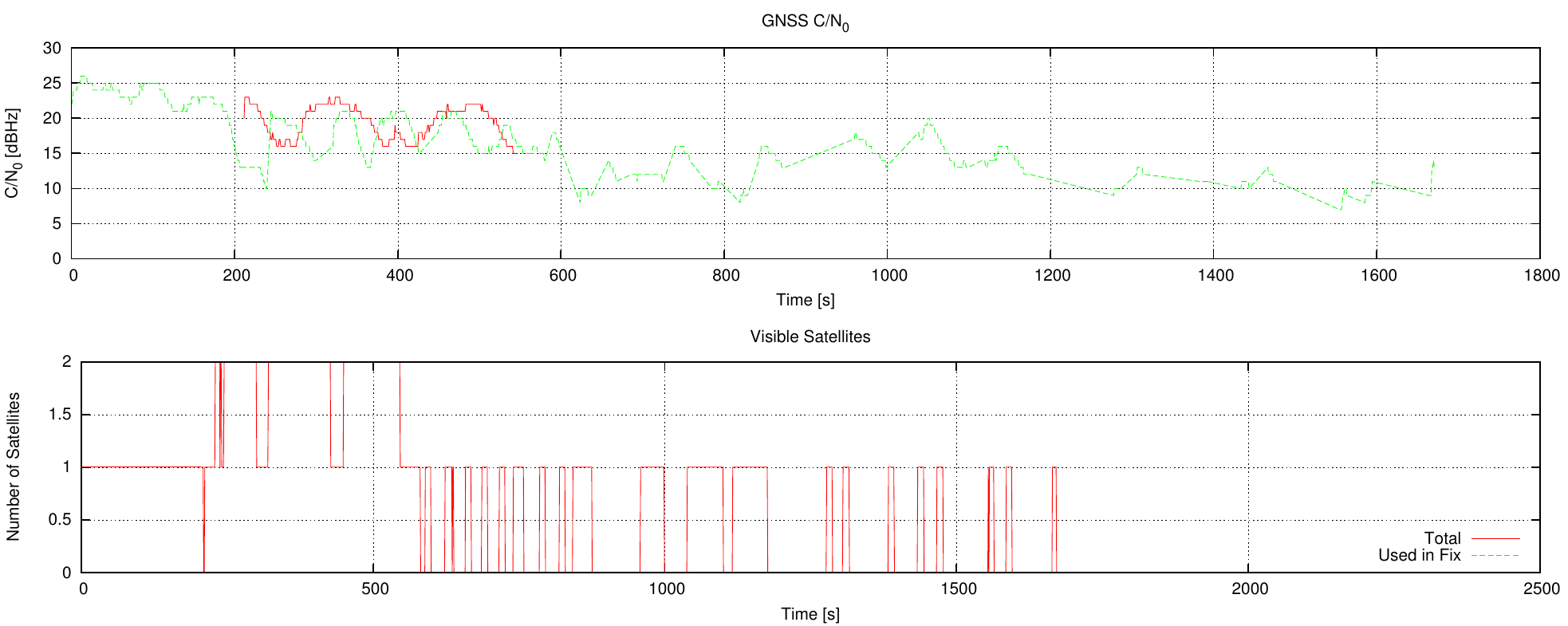}}
\caption{Example measurement at Location 3 (office) indoor with no visibility of the sky. The device is initialised at this location and stays there during the measurement.}
\label{fig:measurement-location-3-only}
\end{figure*}

\begin{figure*}[htbp]
\centerline{\includegraphics[width=.9\linewidth]{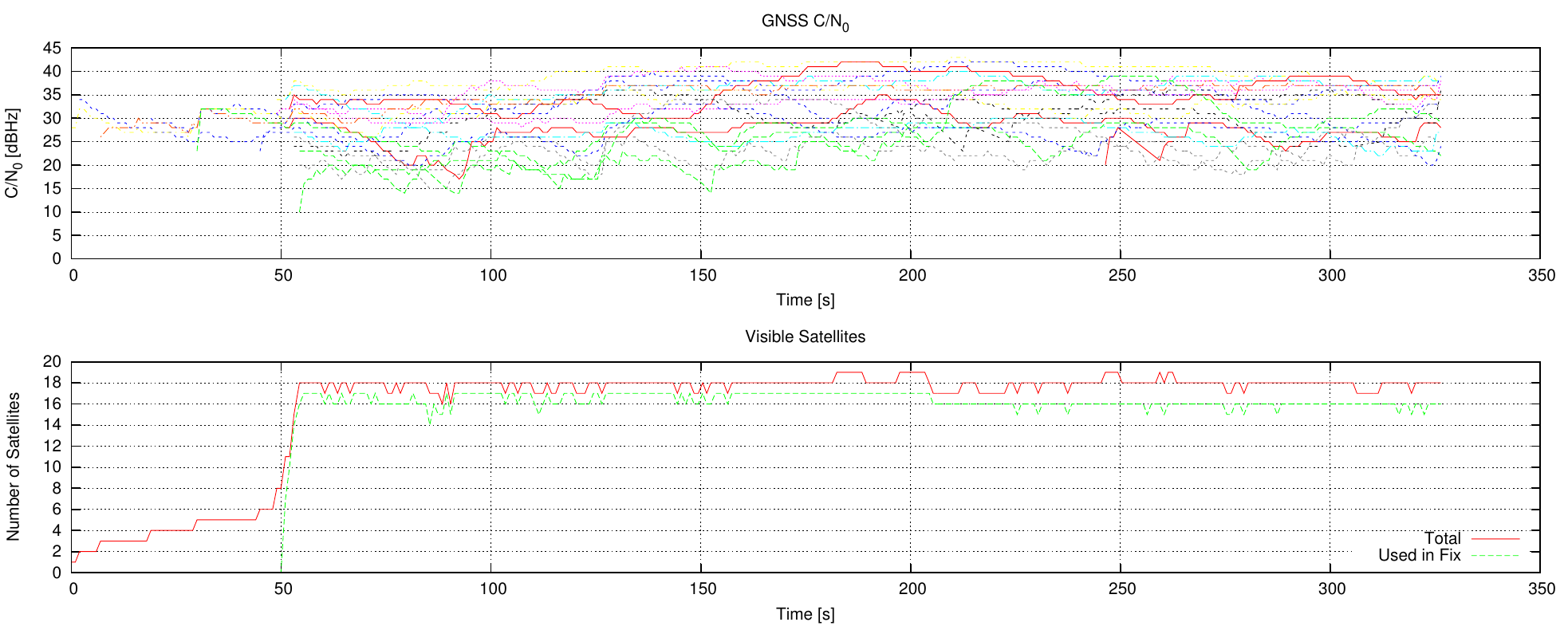}}
\caption{Example measurement at the parking lot location with good visibility of the sky. The device is initialised at this location and stays there during the measurement.}
\label{fig:measurement-location-4-only}
\end{figure*}

Measurements were taken again at these locations separately, i.e. the device was shut down and started up at the location in order to show the initialization phase and the stabilization of measurements.  show the results. While the time to first fix is less than 60\,s in the outdoor cases (Figure~\ref{fig:measurement-location-1-only} and \ref{fig:measurement-location-4-only}), the indoor locations shows a significant waiting time over 300\,s or no fix at all (Figures~\ref{fig:measurement-location-2-only} and \ref{fig:measurement-location-3-only}). Furthermore, in location 2 it is visible that (Figure~\ref{fig:measurement-location-2-only}) the majority of satellites have  $C/N_0$ between 15 and 20\,dB-Hz, while 3 satellites reach a maximum over 25\,dB-Hz. In contrast, on the parking lot (see Figure~\ref{fig:measurement-location-4-only}), the majority of satellites are received with $C/N_0$ over 20\,dB-Hz. 

\begin{figure}[htbp]
\centerline{\includegraphics[width=\linewidth]{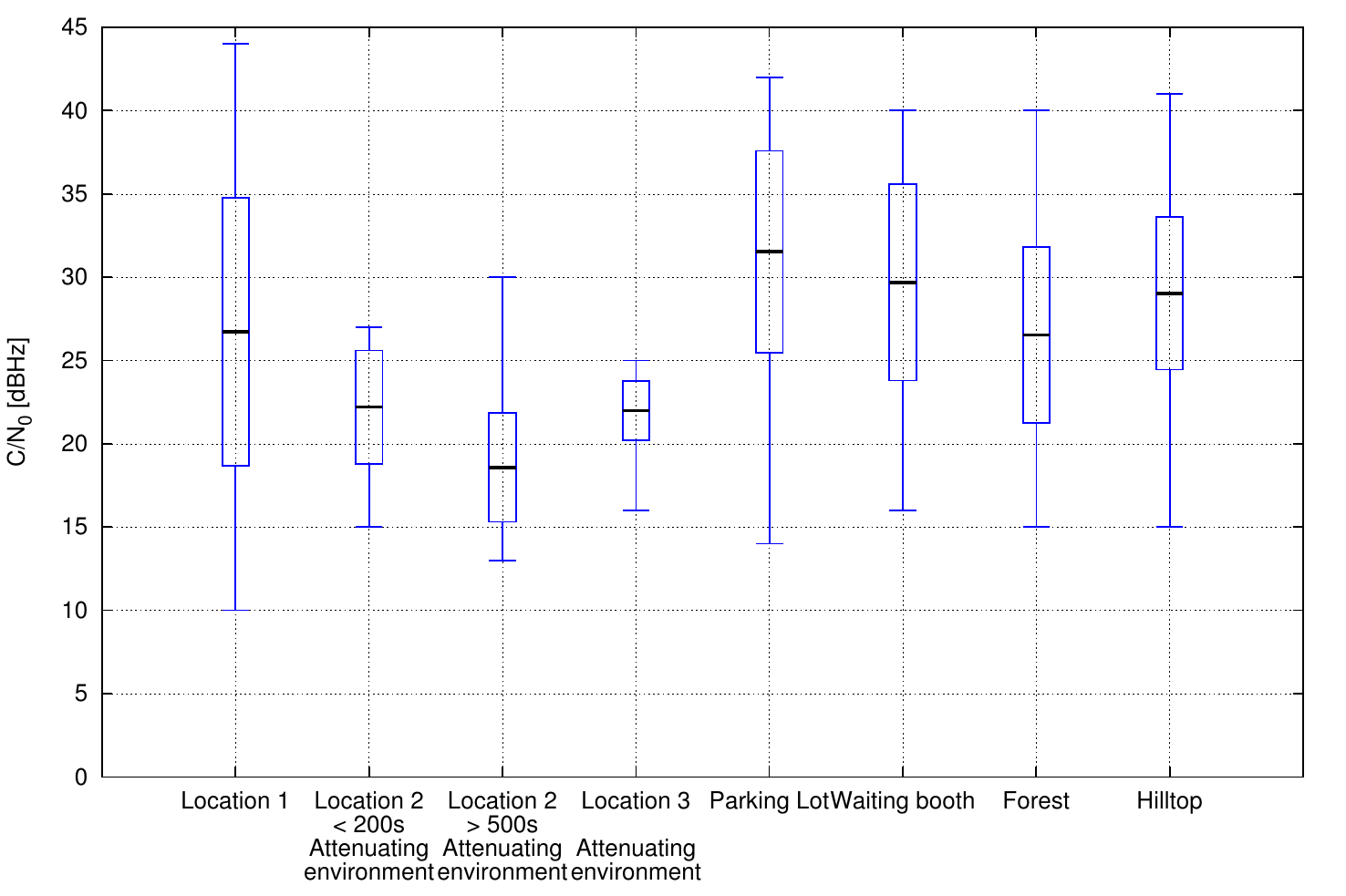}}
\caption{$C/N_0$ statistics for different locations. The box-plot is centered at the sample mean $\mu$ with the box indicating the standard deviation ($\mu \pm \sigma$). Upper and lower whiskers indicate mininum and maximum of the measured values. A measurement duration of 100\,s was used.}
\label{fig:plot-summary-cn0}
\end{figure}

\begin{figure}[htbp]
\centerline{\includegraphics[width=\linewidth]{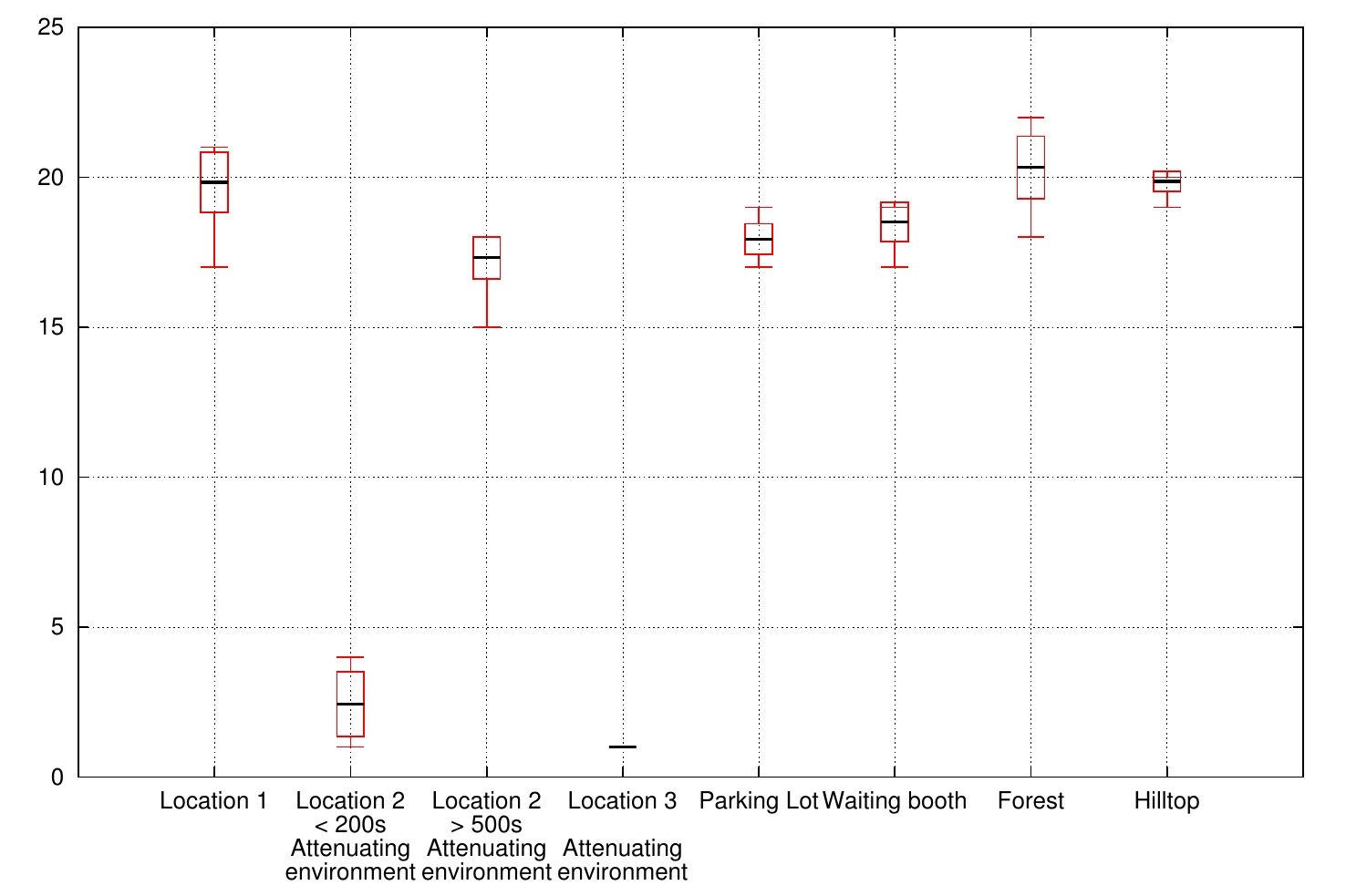}}
\caption{Number of satellite statistics for different locations. The box-plot is centered at the sample mean $\mu$ with the box indicating the standard deviation ($\mu \pm \sigma$). Upper and lower whiskers indicate mininum and maximum of the measured values. A measurement duration of 100\,s was used.}
\label{fig:plot-summary-numsat}
\end{figure}

Rural measurements were taken in a forest and on a hilltop, as well as inside a waiting booth of a railway station, which hass non-tinted glass on all sides (including glass door) and a metal roof. A summary of the measurements are given in Figures~\ref{fig:plot-summary-cn0} and \ref{fig:plot-summary-numsat}.

~
~
~

\section{Detection Approach} \label{sec:detection}

\subsection{Definition of a detection function}

For the detection of an attenuating environment via GNSS we define a function $f_i$ that takes a series of raw data measurements $R$ as input and indicates whether criteria for an attenuating environment are met:
\[ f_i(R) := 
	\left\{ 
	\begin{array}{ll}
		1 & \text{if criteria for an attenuating environment are} \\
		  & \text{met (e.g., $R$ was recorded in a known } \\
		  & \text{attenuating environment)} \\
		0 & \text{otherwise} 
	\end{array} \right. \]

$R$ is defined as a tuple of time series $(c,\alpha,\phi,\rho,\chi)$, which contains for a satellite $k$ at timestamp $i$ the following values:
\begin{itemize}
\item[~] $c_i{(k)}$ is the carrier-over-noise ratio ($C/N_0$). 
\item[~] $\alpha_i{(k)}$ is azimuth of satellite $k$.
\item[~] $\phi_i{(k)}$ is the elevation of satellite $k$.
\item[~] $\rho_i{(k)}$ is $1$ if a signal from $k$ is available, and $0$ otherwise.
\item[~] $\chi_i{(k)}$ is $1$ if satellite $k$ is used in the fix, and $0$ otherwise.
\end{itemize}
Furthermore, let $S_i := \sum_{k} \rho_i(k)$ denote the number of available satellites, 
and $X_i := \sum_{k} \chi_i(k)$ the number of satellites used in a fix at timestamp $i$.

If no special knowledge about the GNSS receiver is known, the function $f_i$ can be specified using an empirical approach. First, empirical data sets $R$ are recorded inside attenuating environments, and data sets $R'$ recorded outside attenuating environments, such as the measurements presented in Section~\ref{sec:results}. Then, criteria to distinguish between $R$ and $R'$ need to be identified. There are several components, where the time series of carrier-to-noise values $c_i^{(k)} $is a candidate we already identified. In cases where this is less obvious, a calculation of statistical distances between empirical distributions could help to identify the relevant component. 

As an example, in the measured carrier-to-noise time series, it is visible that in attenuating environments $\max_k c_i^{(k)} \leq 30$, while this value is significantly exceeded in non-attenuating environments. Thus, this threshold could be the criterion to specify $f_i(R)$. An initialisation phase, e.g. the typical waiting time before the GNSS receiver is able to calculate a fix, should not be taken into account when calculating $f_i$. 

In general, it is also possible to define a function $f_a$ that returns the estimated attenuation value $a$ of the environment:
\[ f_a(R) := a \]
where $a$ could have a continuous or a stepwise value range. In the following we focus on a binary detection function $f_i(R)$.

\subsection{Online detection}
 
For an online detection, we distinguish between initialization or start-up phase of the device and the stable phase of operation.  
This initialisation duration $d_0$ is defined as the initial waiting time duration before the GNSS receiver chipset can deliver measurement values. This could be a time interval for a typical time-to-first-fix, taken from empirical measurements.  
The actual measurement duration $d_m$ is the time duration in which measurements are recored as input for the measurement. 
It is important not to underestimate the duration $d_0$ of the initalizatioin, since the number or values of measured signals could be lower than in later phases and could lead to false positive detection of an attenuating environment.

In general, the following criteria for detecting an attenuating environment could be applied (among others):
\begin{itemize}
	\item Threshold on average $C/N_0$: For all satellites i, the average $C/N_0(i)$ is below a threshold within the measurement duration.
	\item Threshold on maximum $C/N_0$: For all satellites i, the maximum $C/N_0(i)$ is below a threshold within the measurement duration.
	\item Threshold on number of detected satellites: The number of distinct satellites, from which a signal is received, is below a threshold within the measurement duration.
	\item Threshold on number of satellites used in a fix: The number of satellites used in the fix is below a threshold within the measurement duration.
\end{itemize}
Thresholds will depend on the device used, since they depend on the GNSS chipset and further aspects of system integration such as the antenna used.

The first three of the aforementioned criteria are shown in Figures~\ref{fig:plot-summary-cn0} and {fig:plot-summary-numsat}. It is clearly visible that Locations 2 and 3 which are characterised as attenuating environment, show different values than other locations. The number of satellites seems to be a clear indicator, however, there is a quite long initialisation phase in Location 2 until the number of detected satellites becomes quite stable. After an initial phase, the number of satellites in Location 2 become similar to other (non-attenuating) locations, while there an attenuation has been measured in Location 2. Here, the carrier-over-noise density maximum seems to be a better indicator, because it is available when the first satellite is detected, and a relevant observation time seems to be smaller. It should be noted, that $C/N_0$ also depends on elevation \cite{fortunato2019real-time}, and in order to eliminate errors from low elevation satellite signals, the initialisation phase could be extended until satellites with higher elevation are in the sample set. 

For the device used in our tests, the following values can be used for detecting an attenuating environment:
\begin{itemize}
\item Initialisation duration $d_0$ = 100\,s
\item Measurement duration $d_m$ = 100\,s
\item Maximum $C/N_0$ threshold of 30\,dB-Hz \\
	i.e. an attenuating environment is detected if $\forall i \in d_m: \max_k c_i^{(k)} \leq 30$ is observed.
\end{itemize}
Note that these values depend on the model of the GNSS receiver and have only been based on the recorded measurements.

\subsection{Discussion}  \label{sec:discussion}

The proposed approach shows that it is possible to distinguish between attenuating environments and non-attenuating environments. The detection thresholds found for the proof of concept are easy to implement and do not require complex computations. However, they are specific to a certain GNSS receiver model. For different models, the approach to derive a detection function in Section~\ref{sec:detection} needs to be performed. For a generalization, more measurements would be needed. 

There are further aspects to consider:
\begin{itemize}
\item There might be materials where attenuation of GNSS signals at 1-1.6\,GHz is different from attenuation at 5\,GHz. This could lead to false positive or false negative results.
\item The use of pseudolites (ground-based pseudo-satellite transmitters) would lead to a false negative result. Pseudolites, which are recommended to use dedicated codes \cite{eccrep168}, could be omitted from the measurement.
\item Many multi-constellation GNSS chipsets are on the market. When using other constellations, their characteristic need to be taken into account, e.g. QZSS signals can only be recevied in Asia/Oceania.
\item Mobility of the device was not in the focus of this study. In case of mobility, a continuous monitoring needs to be performed in order to detect transitions as in Figure~\ref{fig:measurement-moving}. Furthermore, for battery-powered devices, also energy-efficiency has to be considered as described in \cite{chen2017satprobe}. 
\end{itemize}

The proposed method is expected to work for single devices or those controlling other connected devices (e.g. a WLAN access point that controls other stations). Weak GNSS signals received by a client device do not imply that also the access point is in an attenuating environment. If the access point is outside an attenuating environment, and a client device is inside, then a detection by the client is not meaningful.

\section{Conclusion} \label{sec:conclusion}

We have demonstrated in a proof-of-concept that GNSS raw data measurements can help to automatically detect environments in which radio signals are attenuated. 
Such environment could be use by a radio device (such as 5GHz WLAN) to decide automatically whether certain radio channels are available or not. This could relieve users from questions of correct placement and/or configuration. 

The proof of concept is based on a tablet computer and software interfaces available from Android 7, which are already available for several years, i.e. the method is applicable to established mass market equipment. Measurements have been taken in urban and rural locations -- two of these locations were attenuating environments. GNSS carrier-over-noise measurements could provide an indication within a short time after initialisation.  

For a generalisation, further measurements in other environments and with other devices need to be performed. This could be one aspect of future work. Another aspect could be the quantification of the attenuation, e.g. how the measured attenuation in the GNSS frequency bands translate into attenuation in 5\,GHz WLAN bands, and whether it is possible to detect is with GNSS receivers.


\newpage

\begin{thebibliography}{00}
\bibitem{angskog2015measurement} \"Angskog, P., B\"ackstr\"om, M., Vallhagen, B.: 
Measurement of Radio Signal Propagation through Window Panes and Energy Saving Windows. In: Proceedings of Electromagnetic Compatibility (EMC), 2015 IEEE International Symposium on (pp. 74-79), 2005.
\bibitem{chen2017satprobe} Kongyang Chen, Guang Tan: SatProbe: Low-energy and fast indoor/outdoor detection based on raw GPS processing. IEEE INFOCOM 2017 - IEEE Conference on Computer Communications. https://doi.org/10.1109/INFOCOM.2017.8057095
\bibitem{dejong2014precise} Kees de Jong, Matthew Goode, Xianglin Liu, and Mark Stone: Precise GNSS Positioning in Arctic Regions, Arctic Technology Conference 2014
\bibitem{eccdec0408} ECC Decision of 09 July 2004 on the harmonised use of the 5 GHz frequency bands for the implementation of Wireless Access Systems including Radio Local Area Networks (WAS/RLANs), ECC/DEC/(04)08, 30 October 2009. https://www.ecodocdb.dk/document/381
\bibitem{eccrep168} ECC Report  168. Regulatory Framework for Indoor GNSS Pseudolites, May, 2011. https://www.ecodocdb.dk/document/276
\bibitem{fortunato2019real-time} Marco Fortunato, Michela Ravanelli, Augusto Mazzoni: Real-Time Geophysical Applications with Android GNSS Raw Measurements. Remote Sensing. 2019, 11, 2113; https://dx.doi.org/10.3390/rs11182113
\bibitem{gsa17whitepaper} European GNSS Agency: Using GNSS Raw Measurements on Android Devices, Whitepaper, 2017. 
\bibitem{hein2008gnss} G\"unter Hein, Andreas Teuber, Hans-J\"org Thierfelder, Anne Wolfe: GNSS Indoors - Fighting the Fading, Part 2. InsideGNSS, May/June 2008.
\bibitem{joseph2010gnss} Angelo Joseph: GNSS Solutions: Measuring GNSS Signal Strength, InsideGNSS, November/December 2010.
\bibitem{kjaergaard2010indoor} Mikkel Baun Kj{\ae}rgaard, Henrik Blunck, Torben Godsk, Thomas Toftkj{\ae}r, Dan Lund Christensen, Kaj Gr{\o}nb{\ae}k: Indoor Positioning Using GPS Revisited. Pervasive Computing 2010, Springer Berlin Heidelberg, pp. 38--56.
\bibitem{li2014iodetector} Mo Li, Pengfei Zhou, Yuanqing Zheng, Zhenjiang Li, and Guobin Shen. 2014. IODetector: A generic service for indoor/outdoor detection. ACM Transactions on Sensor Networks 11, 2, Article 28 (December 2014), 29 pages.
DOI: http://dx.doi.org/10.1145/2659466
\bibitem{schmidt08uwb} Ingo Schmidt, Achim Enders, Martin Schwark, Jens Schuur, Robert Geise, Martin Schirrmacher, Henning  St\"ofen: UWB aircraft transfer function measurements in the frequency range from 2 to 8 GHz, 2008 International Symposium on Electromagnetic Compatibility - EMC Europe, Hamburg, 2008, pp. 1-4, https://doi.org/10.1109/EMCEUROPE.2008.4786855
\end{thebibliography}
\end{document}